\documentclass[aps,prb,twocolumn,
superscriptaddress,onlinecite]{revtex4-1}

\usepackage{graphicx}
\usepackage{epstopdf}
\usepackage{siunitx}

\begin{document}


\title{Magnetism and superconductivity in U$_2$Pt$_x$Rh$_{(1-x)}$C$_2$}


\author{N. Wakeham}
\affiliation{Los Alamos National Laboratory}

\author{Ni Ni}%
\affiliation{Department of Physics and Astronomy, UCLA}%

\author{E. D. Bauer}
\affiliation{Los Alamos National Laboratory}

\author{J. D. Thompson}
\affiliation{Los Alamos National Laboratory}

\author{E. Tegtmeier}
\affiliation{Los Alamos National Laboratory}

\author{F. Ronning}
\affiliation{Los Alamos National Laboratory}


\date{\today}

\begin{abstract}
We report the phase diagram of the doping series U$_2$Pt$_x$Rh$_{(1-x)}$C$_2$, studied through measurements of resistivity, specific heat and magnetic susceptibility. The N\'{e}el temperature of U$_2$RhC$_2$ of $\sim 22$ K is suppressed with increasing Pt content, reaching zero temperature  close to $x=0.7$, where we observed signatures of increased quantum fluctuations. In addition, evidence is presented that the antiferromagnetic state undergoes a spin-reorientation transition upon application of an applied magnetic field. This transition shows non-monotonic behaviour as a function of $x$, peaking at around $x=0.3$. Superconductivity is observed for $x\geq0.9$, with $T_{\rm{c}}$ increasing with increasing $x$. The reduction in $T_{\rm{c}}$ and increase in residual resistivity with decreasing Pt content is inconsistent with the extension of the Abrikosov-Gor'kov theory to unconventional superconductivity.
\end{abstract}


\maketitle
\section{Introduction}
The study of superconductivity and magnetism in uranium based heavy-fermion materials has provided a rich array of physics that has challenged and guided our understanding for several decades. Notable examples are the superconductivity in UBe$_{13}$ \cite{Ott1983}, as well as the antiferromagnetic (AFM) order, multiple superconducting phases and evidence for triplet pairing seen in  UPt$_3$ \cite{RevModPhys.74.235}. Recently, the study of unconventional superconductivity has focussed on the role of quantum fluctuations associated with the suppression of a second order phase transition to zero temperature, such as in URhGe under an applied magnetic field \cite{Aoki2001,Levy2001}. U$_2$PtC$_2$ is a so-called ``nearly-heavy-fermion'' system because of its moderately enhanced electron effective mass of order 100 times that of a free electron \cite{Meisner1984}. It becomes superconducting below the transition temperature $T_{\rm{c}} = 1.47$\,K, and does not display long range magnetic order\cite{Matthias1969,Meisner1984,Wu1994a,Ebel1996}. The mechanism for superconductivity in U$_2$PtC$_2$ is an open question. Recent nuclear magnetic resonance (NMR) measurements of U$_2$PtC$_2$ have shown evidence for spin triplet pairing and unconventional superconductivity \cite{Mounce2014}. Isostructural U$_2$RhC$_2$ displays no superconductivity but orders antiferromagnetically at a N\'{e}el temperature $T_{\rm{N}}\sim 22$\,K, and shows evidence for possible ``complex magnetic behaviour" at lower temperatures \cite{Ebel1996}.

Given the nature of the parent compounds, the doping series U$_2$Pt$_x$Rh$_{(1-x)}$C$_2$ must show some evolution between a magnetically ordered and superconducting ground state. Here we report thermodynamic and transport measurements of U$_2$Pt$_x$Rh$_{(1-x)}$C$_2$ and find that with increasing platinum content $T_{\rm{N}}$ is suppressed to zero temperature close to $x=0.7$, where we observe evidence for quantum critical fluctuations. However, superconductivity is not observed until $x=0.9$, where signatures of these fluctuations are almost entirely absent. $T_{\rm{c}}$ is maximal in U$_2$PtC$_2$. Study of the magnetic field dependence of the AFM state has revealed evidence of a spin-reorientation transition. We discuss the possible implications of our results with respect to competing magnetic interactions and superconductivity.

\section{Sample preparation}
\begin{figure}[h]
 \includegraphics[width=1\columnwidth]{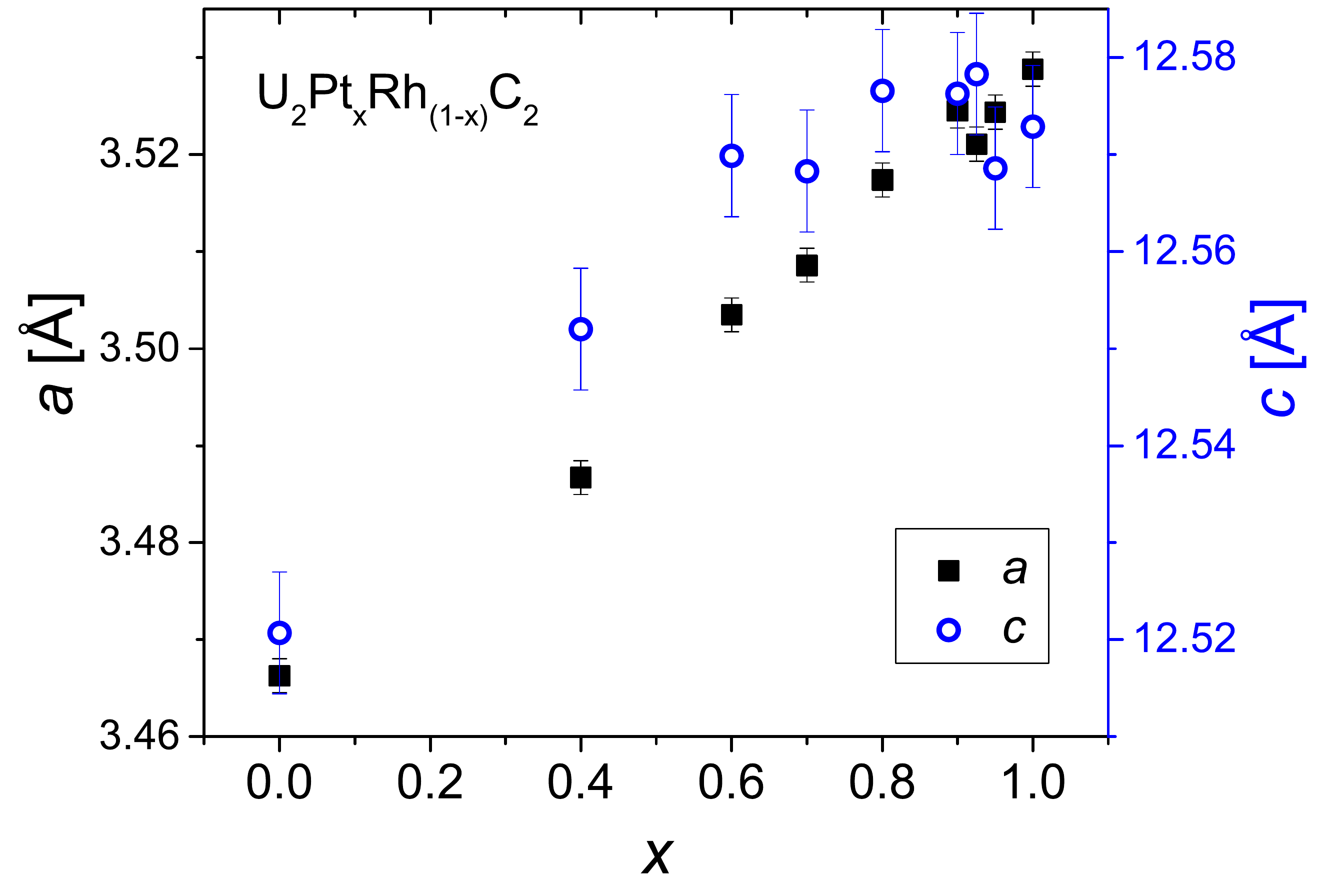}%
\caption{(Color online) Lattice parameters of U$_2$Pt$_x$Rh$_{(1-x)}$C$_2$ as a function of $x$, determined by powder X-ray diffraction. \label{LatticeParam}}
\end{figure}

Polycrystalline U$_2$Pt$_x$Rh$_{(1-x)}$C$_2$ samples were made by arc melting and subsequent annealing. Depleted-U, Pt, Rh and C were weighed out according to the ratio of U:Pt:Rh:C of $2: 1.1x:1.1(1-x):2.2$. This ratio was found to produce a lower UC impurity content than the other ratios attempted of $2: x:(1-x):2$ or $2: 2x:2(1-x):2.4$. During the arc melting, the resulting button was flipped and melted several times. The button was then wrapped in Ta foil and sealed in a quartz tube under vacuum before being annealed for two months at 1050$^{\circ}$C. Annealing of the product was necessary to further reduce the impurity content. Growths of single crystals were attempted using Bi, Zn, Al, Ga, Sn, Sb, and U fluxes, but were not successful. Powder X-ray diffraction of the polycrystalline samples confirmed that U$_2$PtC$_2$ and U$_2$RhC$_2$ possess the Na$_2$HgO$_2$ structure type in which all the U sites are equivalent, as first discussed for the isostructural case of U$_2$IrC$_2$ \cite{Bowman1971}. The lattice parameters as a function of $x$ are shown in Fig. \ref{LatticeParam}.

\begin{figure}[h]
 \includegraphics[width=1\columnwidth]{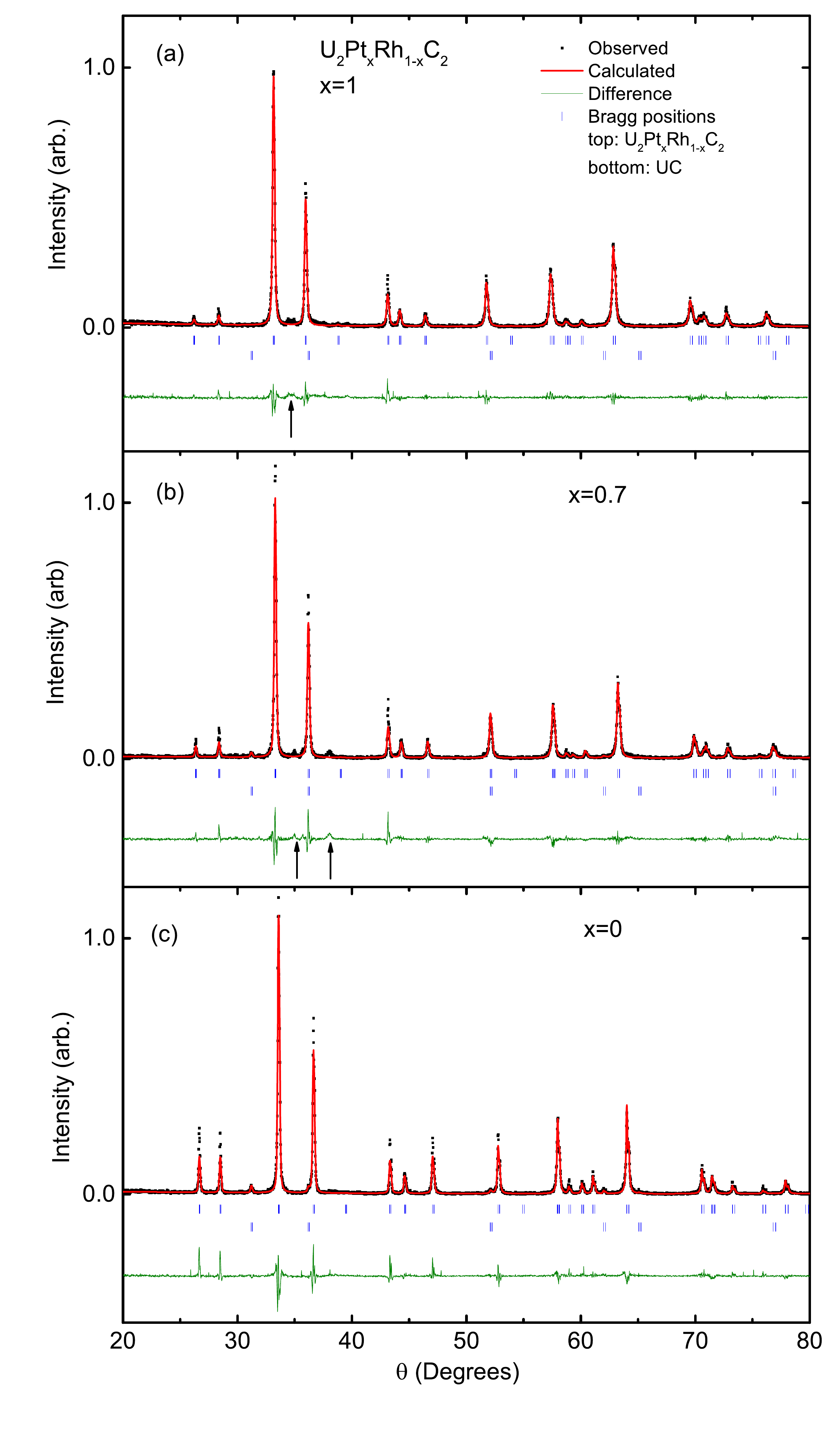}%
 \caption{(Color online) Observed and calculated powder X-ray diffraction pattern for (a) U$_2$PtC$_2$ (b) U$_2$Pt$_{0.7}$Rh$_{0.3}$C$_2$, and (c) U$_2$RhC$_2$. Upper blue marks indicate Bragg positions for U$_2$Pt$_x$Rh$_{(1-x)}$C$_2$, lower blue marks indicate Bragg positions for UC. Black arrows highlight peaks originating from an unknown phase.\label{xrayBoth}}
 \end{figure}

\begin{figure}[h]
 \includegraphics[width=1\columnwidth]{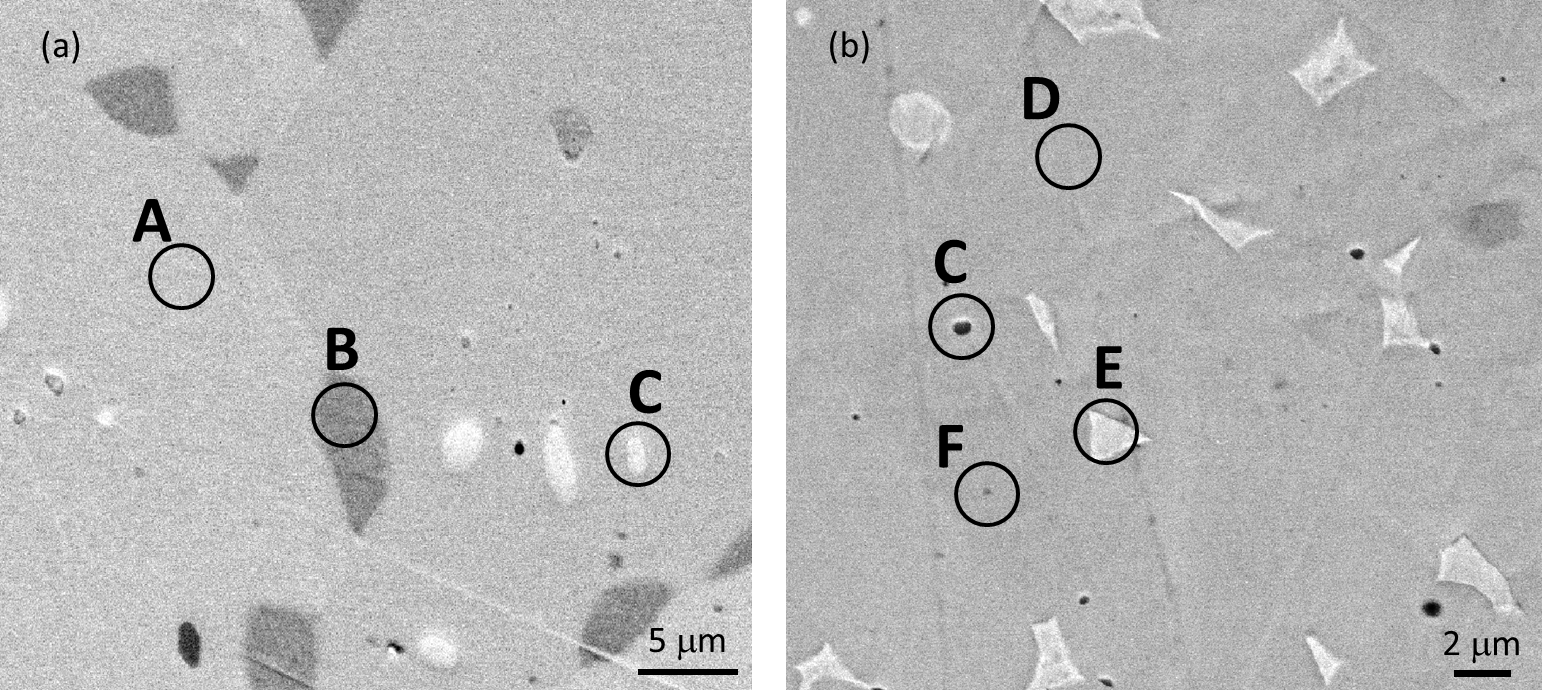}%
 \caption{Backscattered electron image of (a) U$_2$RhC$_2$ and (b) U$_2$PtC$_2$. Labelled circles highlight different material phases, which are described in the main text.\label{EDX}}
 \end{figure}

The powder X-ray diffraction patterns for U$_2$PtC$_2$, U$_2$Pt$_{0.7}$Rh$_{0.3}$C$_2$ and U$_2$RhC$_2$  are shown in Fig. \ref{xrayBoth}a, \ref{xrayBoth}b and \ref{xrayBoth}c, respectively.
Detectable impurity phases in powder X-ray patterns of the various doped samples were the paramagnetic materials UC, UC$_2$, Rh or Pt. These were all at a concentration of less than $10\%$. In samples with $x>0.2$ there was also a small number of low intensity peaks that could not be identified. In order to further investigate the sample quality, energy dispersive X-ray (EDX) measurements were performed on U$_2$PtC$_2$ and U$_2$RhC$_2$. Backscattered electron images of these two samples are shown in Fig. \ref{EDX}.
 Fig. \ref{EDX}a shows an image of the U$_2$RhC$_2$ sample in which we identify three phases, (A) U$_2$RhC$_2$, (B) URh$_2$C$_y$ ($\sim 5\%$) and (C) UC ($\sim 1\%$). Fig. \ref{EDX}b shows an image of the U$_2$PtC$_2$ sample in which we also identify three phases, (C) UC($< 0.1\%$), (D) U$_2$PtC$_2$ and (E) UPt$_2$C$_y$ ($\sim 5\%$). The concentrations of the impurity phases were estimated from the area of the features in the images. The concentrations of UC calculated from the Rietveld refinement of the x-ray diffraction patterns shown in Fig. \ref{xrayBoth} are $0\%$ and $1.9\%$ for the U$_2$PtC$_2$ and U$_2$RhC$_2$ samples, respectively. The region marked F in Fig. \ref{EDX}b indicates a small inclusion that could not be identified because the diameter of the feature is significantly smaller than the X-ray spot size, which is of order \SI{1}{\micro\metre}. The exact carbon content of the UPt$_2$C$_y$ and URh$_2$C$_y$ phases could not be established within the measurement because of the difficulty in resolving the concentration of light elements. However, we were able to estimate that $y\leq 2$. The powder X-ray diffraction data did not indicate the presence of a UPt$_2$C$_y$ or a URh$_2$C$_y$ phase, although there were unidentified peaks in the U$_2$PtC$_2$ sample. Therefore, we conclude that either the UPt$_2$C$_y$ is responsible for the unidentified X-ray peaks, or that the phase is amorphous and not visible in the X-ray pattern. If the unidentified peaks are not the result of the UPt$_2$C$_y$ inclusions, then we must assume that these come from the small features marked as F in the electron image. No unidentified peaks were seen in the X-ray of the U$_2$RhC$_2$ samples, despite observation of URh$_2$C$_y$ inclusions from the EDX measurements. This suggests that these inclusions are amorphous. 

\section{Phase diagram}\label{Phase}
\begin{figure}[h]
 \includegraphics[width=1\columnwidth]{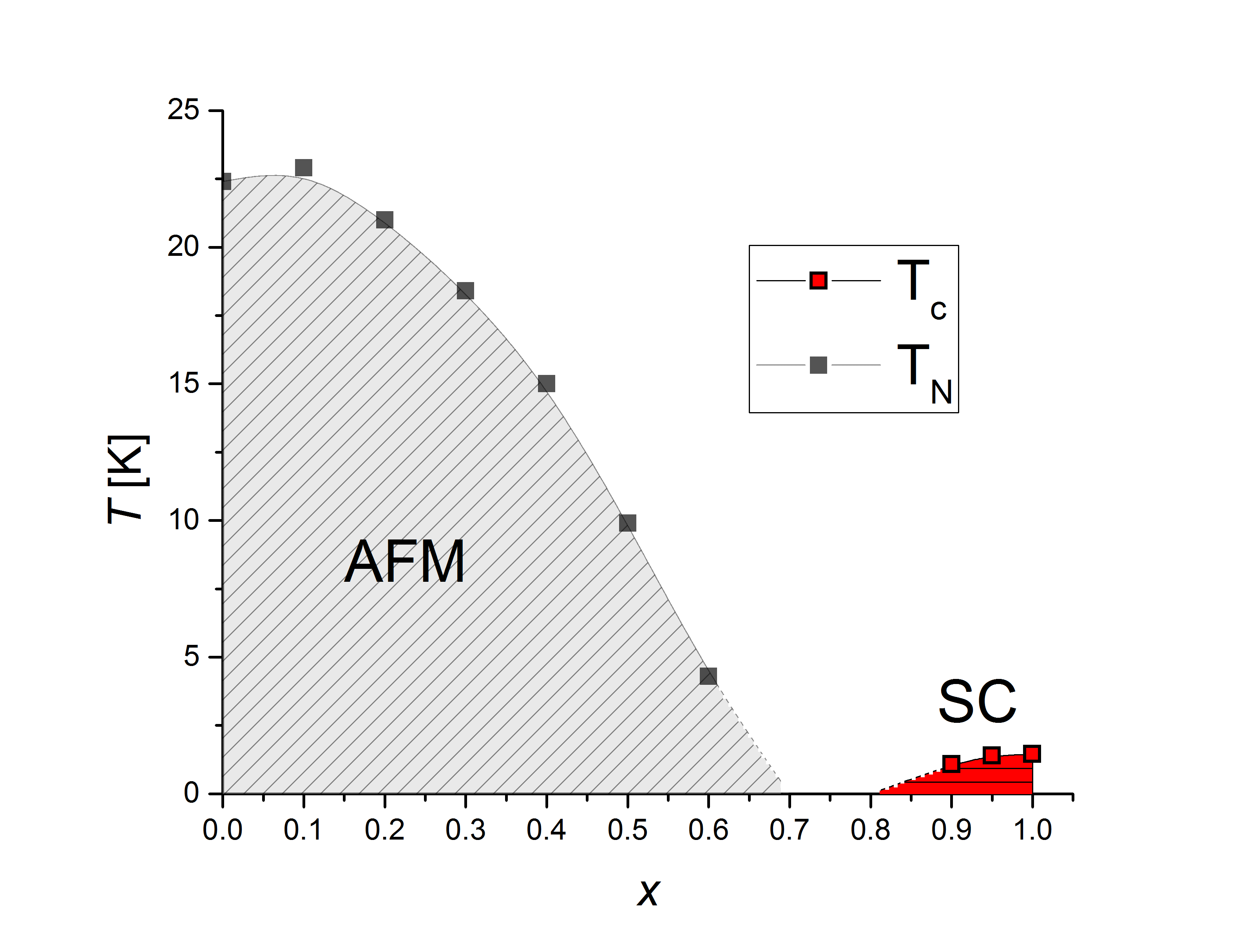}%
 \caption{(Color online) Phase diagram of U$_2$Pt$_x$Rh$_{(1-x)}$C$_2$ as a function of $x$, as determined from specific heat, magnetic susceptibility and resistivity measurements. Gray region indicates the antiferromagnetic (AFM) state with transition temperature $T_N$, red region indicates the superconducting (SC) state with transition temperature $T_c$. Lines are guides to the eye. \label{PhaseDiag}}
 \end{figure}

\begin{figure}[h]
 \includegraphics[width=0.95\columnwidth]{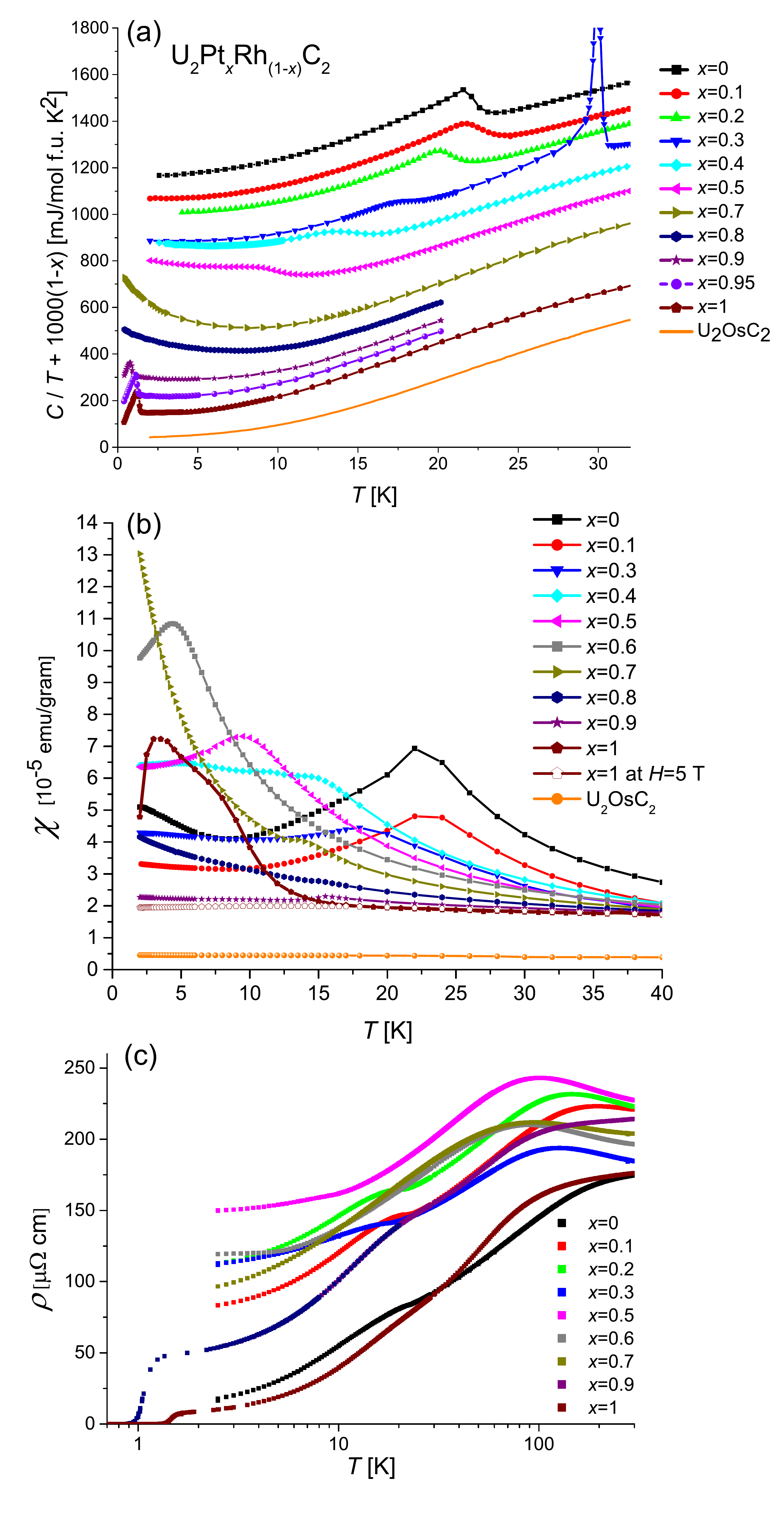}
 \caption{(Color online) Specific heat, magnetic susceptibility and resistivity of U$_2$Pt$_x$Rh$_{(1-x)}$C$_2$ samples. (a) Temperature dependence of the specific heat divided by temperature of U$_2$Pt$_x$Rh$_{(1-x)}$C$_2$, offset by a factor of $1000(1-x)$ mJ/mol\,f.u.\,K$^2$. The specific heat of the isostructural paramagnetic analogue U$_2$OsC$_2$ is also included with zero offset. (b) Temperature dependence of the magnetic susceptibility of U$_2$Pt$_x$Rh$_{(1-x)}$C$_2$ and U$_2$OsC$_2$. Susceptibility measurements were performed at 0.1\,T unless otherwise stated. (c) Resistivity of U$_2$Pt$_x$Rh$_{(1-x)}$C$_2$ as a function of temperature, shown on a logarithmic temperature scale.\label{C_TandChi}}
 \end{figure}

The zero field phase diagram of U$_2$Pt$_x$Rh$_{(1-x)}$C$_2$ shown in Fig.\ref{PhaseDiag}
was established from measurements of the resistivity $\rho$, specific heat $C$ and magnetic susceptibility $\chi$, shown in Fig. \ref{C_TandChi} as a function of temperature $T$.
 The $C/T$ data shown in Fig. \ref{C_TandChi}a are shifted by $1000(1-x)$ mJ/mol f.u. K$^2$ for clarity. Electrical resistivity of the samples was measured using a four-probe low frequency AC resistance bridge with spot-welded contacts of platinum wires. Specific heat measurements were performed using the time-relaxation method. Both of these measurements were performed within the Quantum Design PPMS system. Magnetic susceptibility was measured using a SQUID magnetometer in the Quantum Design MPMS system.

The AFM transition is observed in U$_2$RhC$_2$ as an anomaly in the specific heat and a peak in the magnetic susceptibility, coincident in temperature, at 21.6\,K. The temperature of the transition slightly increases in $x=0.1$, and then is suppressed and broadened with increasing Pt content. As the N\'{e}el temperature is suppressed to zero temperature with increasing $x$ there is an upturn in $C/T$ at low temperatures, suggestive of quantum fluctuations. The rate of the low temperature increase in $C/T$ is greatest in the $x=0.7$ sample, and at this doping the magnetic susceptibility also increases rapidly at low temperatures. $x=0.7$ is also the lowest doping in which an AFM transition is not observed above 0.38\,K. Superconductivity is observed, above the lowest measured temperature of 0.38\,K, only in samples in which $x\geq0.9$, with T$_c$ increasing with increasing Pt content. At $x=0.3$ a first order anomaly was observed in the specific heat at 30\,K. This anomaly was reproducible between several crystals at this doping, but no corresponding feature was observed in the magnetic susceptibility or resistivity, and this type of anomaly was not observed at any other doping. The origin of the anomaly is not known.

An additional anomaly in the magnetic susceptibility is observed at around 16\,K in some samples with $x\geq0.7$. This is particularly prominent in the $x=1$ measurement at 0.1\,T, although it is not present in the high field data. No corresponding feature is seen in the resistivity or specific heat measurements. These susceptibility features are likely to be an extrinsic impurity contribution. UPt$_2$C$_{0.1}$ and UPt$_2$C$_{0.2}$ were synthesized in order to investigate the possible contribution of the UPt$_2$C$_y$ ($y\leq 2$) impurity seen in EDX measurements. Measurements of the magnetic susceptibility and heat capacity, not shown here, showed a small anomaly in magnetic susceptibility at $\sim 15$\,K, with no corresponding feature in the specific heat. Therefore it is possible that this feature is responsible for the observed anomaly in the U$_2$Pt$_x$Rh$_{(1-x)}$C$_2$ data. In addition, the presence of a small concentration of UPt, which shows a peak in magnetic susceptibility at $\sim 17$\,K, cannot be excluded \cite{Huber1975,Proke}. However, we cannot rule out that the feature may be intrinsic, reminiscent of UPt$_{3}$ \cite{RevModPhys.74.235}.

\section{Magnetic field study}
\begin{figure}[h]
 \includegraphics[width=1\columnwidth]{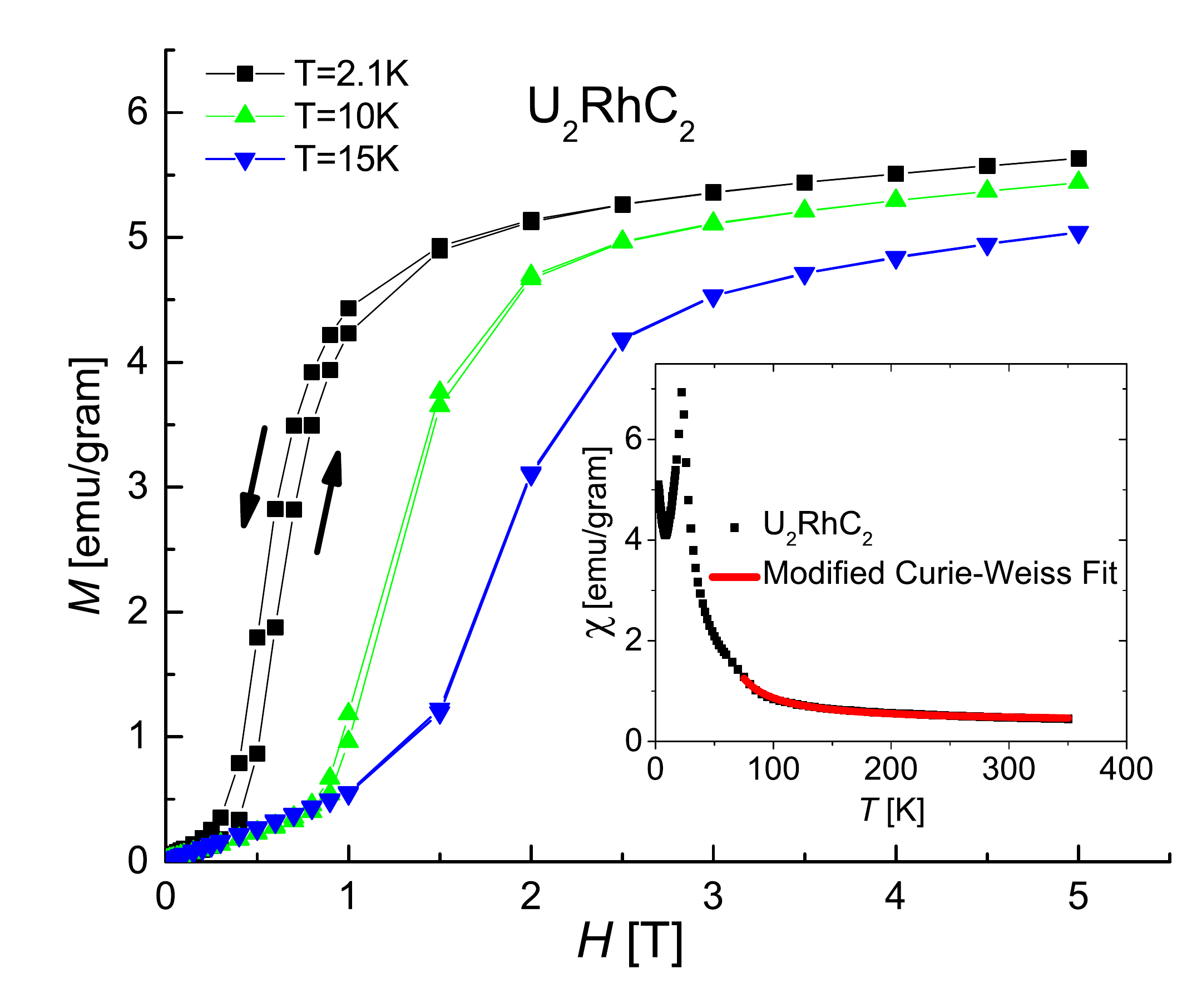}%
 \caption{(Color online) Magnetization $M$ of U$_2$RhC$_2$ as a function of temperature in various applied magnetic fields. Arrows indicate the direction of the field sweep for data at $T=$2.1\,K. Inset: Magnetic susceptibility of U$_2$RhC$_2$ as a function of temperature. Solid red line shows the fit to a modified Curie-Weiss law above 75\,K.  \label{isotherms}}
 \end{figure}

In order to further investigate the magnetism in U$_2$Pt$_x$Rh$_{(1-x)}$C$_2$, thermodynamic and transport measurements were also taken under an applied magnetic field $H$. Isotherms of magnetization $M$ as a function of applied field are shown in Fig. \ref{isotherms} for U$_2$RhC$_2$.
These data show a metamagnetic transition from the AFM state at a characteristic field that increases with increasing temperature. This is likely to be a spin-reorientation transition, such as a spin-flop or spin-flip transition. The transition is clearly first order, as evidenced by the hysteresis observed in $M(H)$ between increasing and decreasing field sweeps. The saturation value of the magnetic moment at 2.1\,K estimated from these data is $\sim 0.3\mu_B/$U. This small value is suggestive of itinerant antiferromagnetism. The inset to Fig. \ref{isotherms} shows the measured magnetic susceptibility as a function of temperature. The data above 75\,K were fitted with a modified Curie-Weiss law because of a large temperature independent contribution. This is given by $\chi=C_c/(T-\Theta)+\chi_0$, where $C_c=$\SI{2.8e-4}\,emu/gram\,K is the Curie constant, $\Theta=44$\,K is the Weiss constant and $\chi_0=$\SI{3.7e-6}\,emu/gram is the temperature independent susceptibility contribution. The effective moment estimated from the Curie constant is $2.8\,\mu_B/$U. This means the Rhodes-Wolfarth ratio of the effective to saturated moment for U$_2$RhC$_2$ is $\sim 9$, which again implies that the magnetism is itinerant \cite{Rhodes1963}.

\begin{figure}[h]
 \includegraphics[width=1\columnwidth]{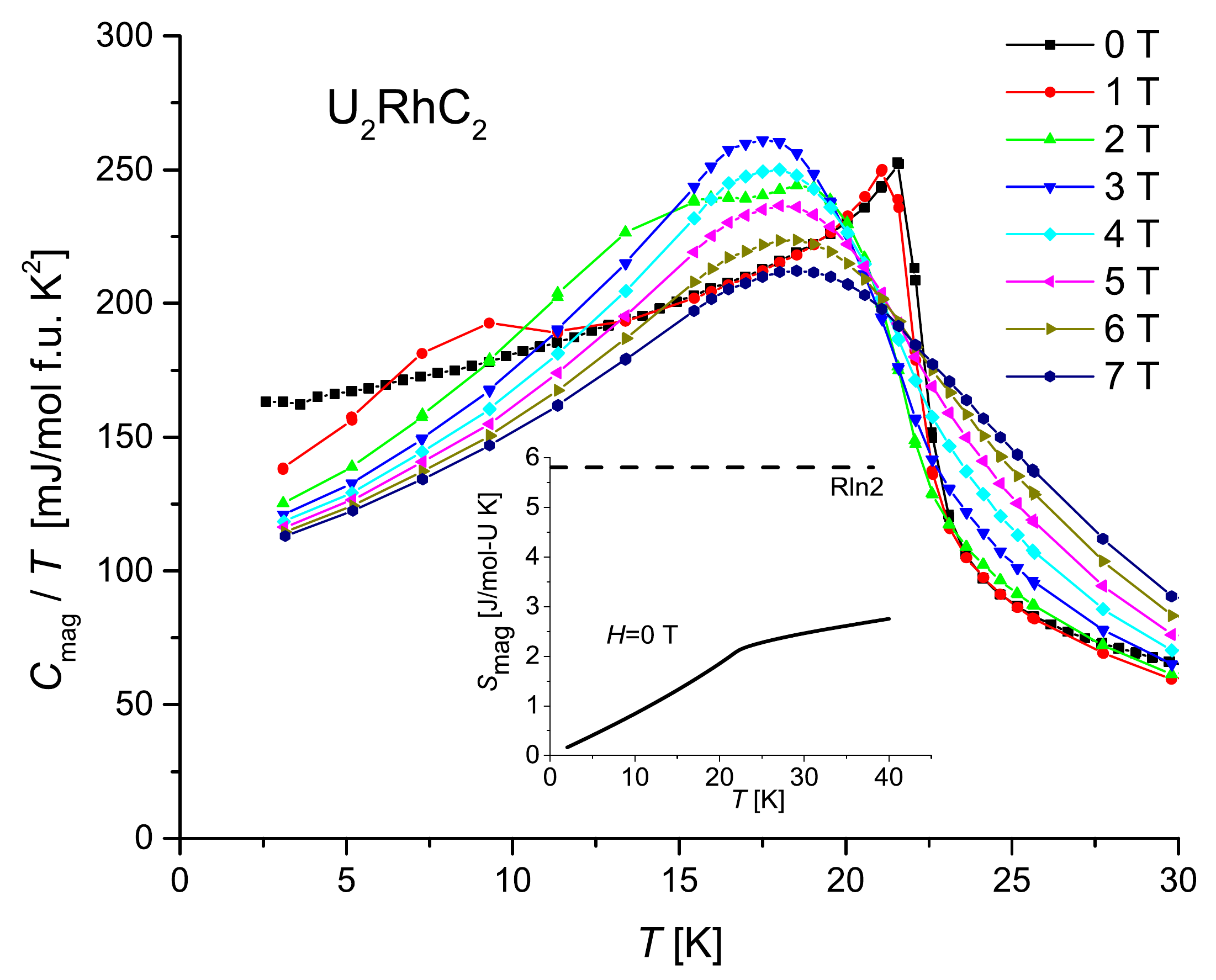}%
 \caption{(Color online) Magnetic contribution to the specific heat of U$_2$RhC$_2$ divided by temperature $C_{\rm{mag}}/T$, as a function of temperature in various applied magnetic fields. Inset: the magnetic entropy $S_{\rm{mag}}$ as a function of temperature. \label{Cmag}}
 \end{figure}

The magnetic contribution to the specific heat $C_{\rm{mag}}$ was isolated from $C$ by subtracting the phonon contribution to the specific heat of the paramagnetic isostructural analogue U$_2$OsC$_2$. The data for $C_{\rm{mag}}/T$ are shown in Fig. \ref{Cmag}.
This figure shows the N\'{e}el temperature in zero field as a sharp peak in $C_{\rm{mag}}/T$ at 21.6\,K. This peak broadens with increasing field, and the peak shifts to lower temperature. At 1\,T there is an anomaly in $C_{\rm{mag}}/T$ near 10\,K, in addition to the AFM transition, which we identify as associated with the spin-reorientation transition. The temperature of this feature increases with increasing field, eventually coinciding with the broadened higher temperature peak. The inset to Fig. \ref{Cmag} shows the magnetic contribution to the entropy as function of temperature, which reaches $\sim 0.4 R \ln(2)$ at $T_N$. The small entropy is also suggestive of itinerant magnetism, and is comparable to other U based antiferromagnets thought to be itinerant, such as UCr$_2$Si$_2$ \cite{D.Matsuda2003} and UPd$_2$Al$_3$ \cite{Geibel1991}. The magnetic anisotropy required to produce a spin-reorientation transition in an itinerant system could be the result of dipole-dipole interactions or spin-orbit coupling. Such scenarios have been suggested in UCr$_2$Si$_2$ \cite{D.Matsuda2003} and (TMTSF)$_2$AsF$_6$ \cite{Mortensen1982}. In zero field the electron specific heat coefficient $\gamma = 160$\, mJ/mol\,f.u.\,K$^2$. Many uranium based materials, such as U$_2$Zn$_{17}$ \cite{Ott1984}, show a significant reduction of $\gamma$ upon entering the magnetically ordered state, and therefore it is likely that the density of states in U$_2$RhC$_2$ in the paramagnetic state is considerably larger than in U$_2$PtC$_2$.

\begin{figure}[h]
 \includegraphics[width=1\columnwidth]{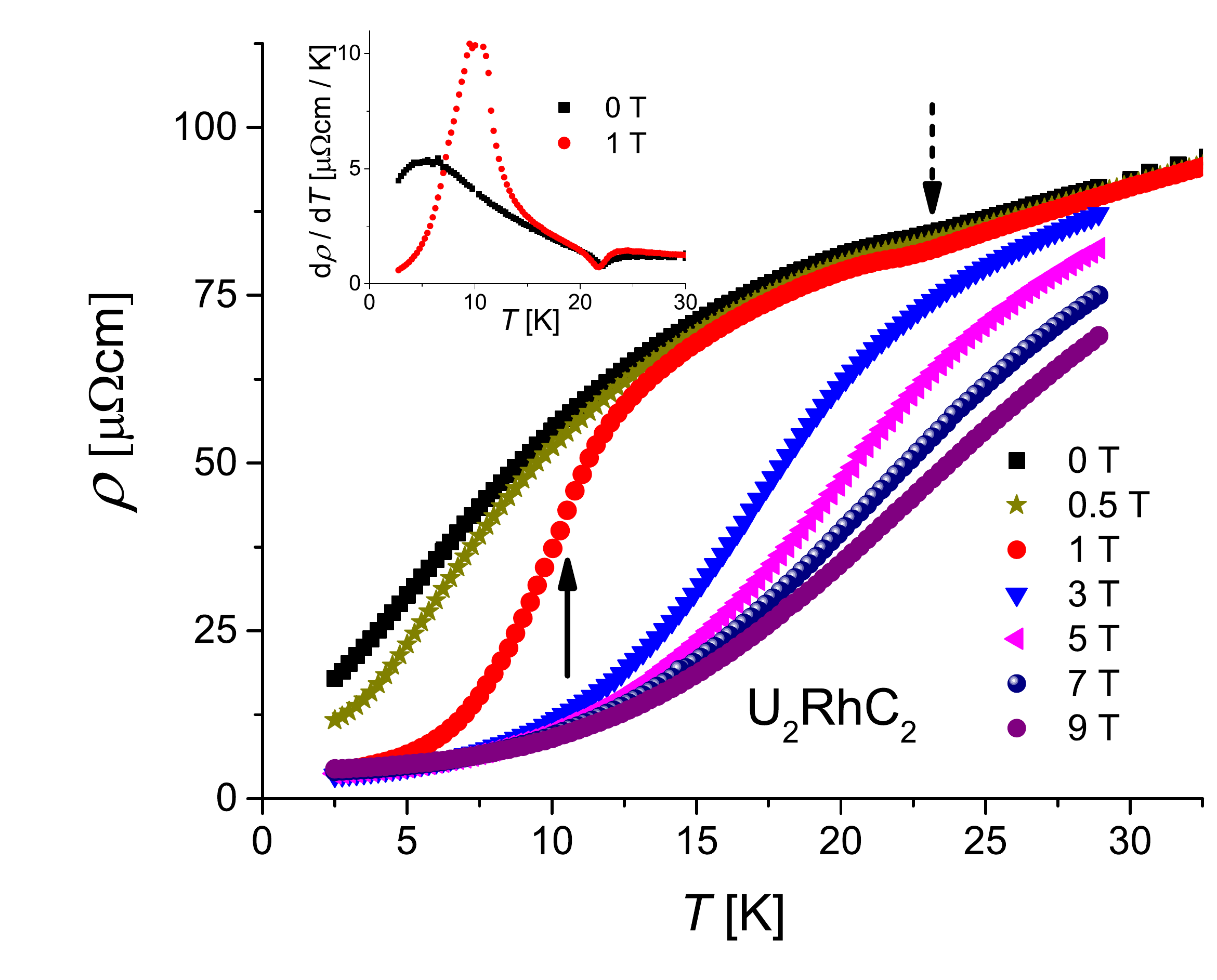}%
 \caption{(Color online) Electrical resistivity $\rho$ of U$_2$RhC$_2$ as a function of temperature in various applied magnetic fields. Dashed arrow indicates the N\'{e}el temperature, solid arrow indicates the spin-reorientation transition. Inset: $\mathrm{d}\rho/\mathrm{d}T$ of U$_2$RhC$_2$ as a function of temperature at 0\,T and 1\,T. \label{rhopt0}}
 \end{figure}

The AFM and spin-reorientation transitions can be seen in $\rho(T)$ at various magnetic fields applied perpendicular to the current, shown in Fig. \ref{rhopt0}.
The N\'{e}el temperature appears as a kink in the zero field $\rho(T)$ curve, as indicated by the dashed arrow, and appears as a dip in $\mathrm{d}\rho/\mathrm{d}T$ shown in the inset to Fig. \ref{rhopt0} for 0\,T and 1\,T. This is reminiscent of the resistivity feature in Cr at $T_N$, where the antiferromagnetism is believed to be itinerant and the kink arises from an energy gap forming on regions of the Fermi surface, leading to a loss of charge carriers and increased resistivity \cite{Reed1967}. However,  in Cr the magnetoresistance is positive, arising from the additional scattering associated with cyclotron orbits of the electrons around the Fermi surface \cite{Reed1967}. In contrast, as shown in Fig. \ref{rhopt0}, the magnetoresistance in U$_2$RhC$_2$ is negative both above and below $T_N$. UNiAl and UNiGa both show anomalies in the resistivity at $T_N$, and display a strong negative magnetoresistance that saturates at high field, in close similarity to the data in Fig. \ref{rhopt0}. The negative magnetoresistance in UNiAl and UNiGa is attributed to field-induced superzone reconstructions of the Fermi surface \cite{Antonov1996,Sechovsky1992}. A similar mechanism may contribute to the large negative magnetoresistance in U$_2$RhC$_2$. In addition, however, the magnetoresistance may have a significant contribution from spin-disorder scattering in the system, and the suppression of the spin fluctuations with magnetic field, as seen, for example, in ferromagnetic (Ga,Mn)As \cite{Matsukura1998}.

\begin{figure}[h]
 \includegraphics[width=1\columnwidth]{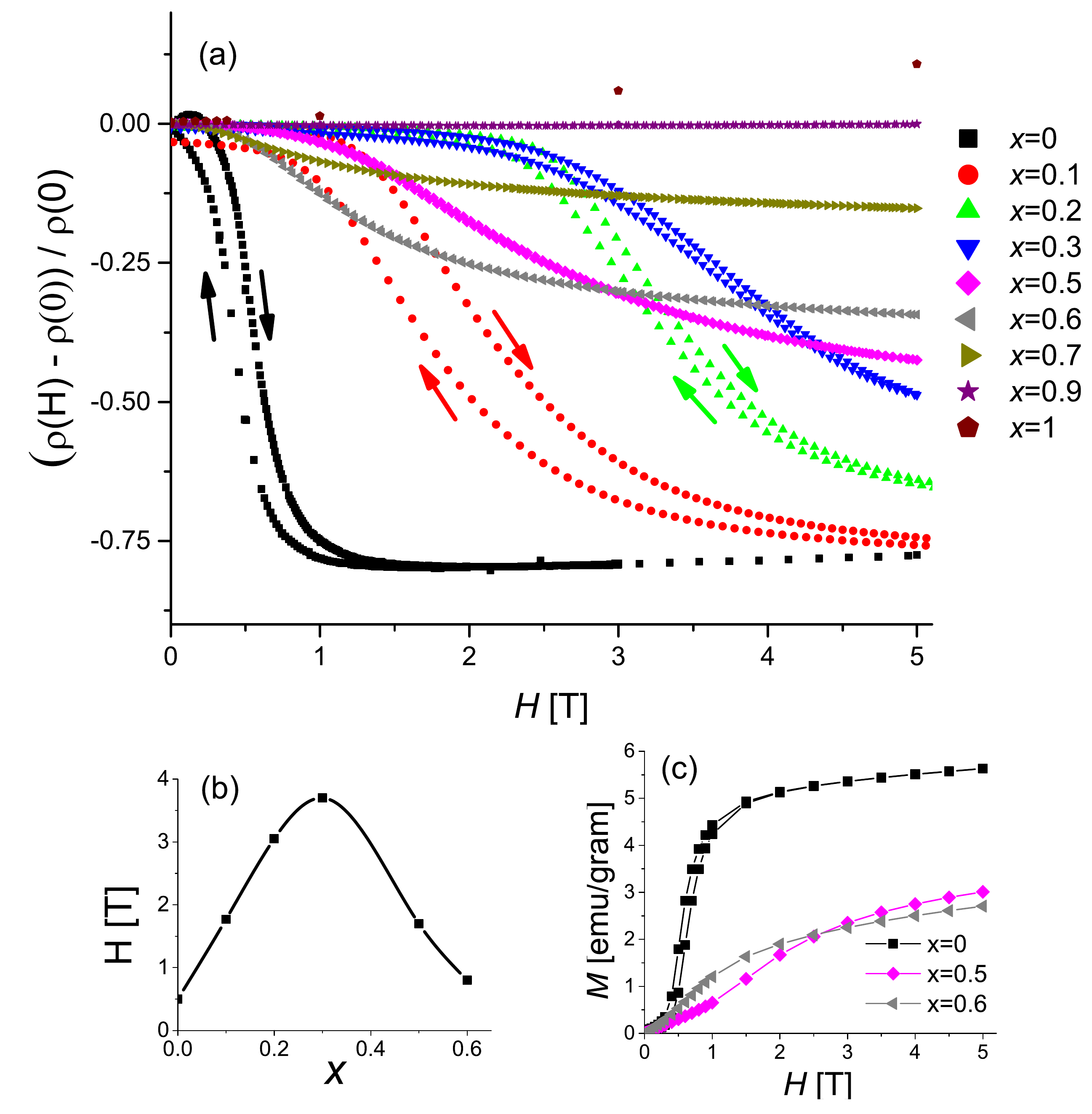}%
 \caption{(Color online) Field dependent features in U$_2$Pt$_x$Rh$_{(1-x)}$C$_2$. (a) Transverse magnetoresistance of U$_2$Pt$_x$Rh$_{(1-x)}$C$_2$ at 2.5\,K. (b) Field of spin-reorientation transition as a function of doping. (c) Magnetization as a function of field for various dopings at 2.1\,K. \label{H_Pt_2K_PhaseDiag}}
 \end{figure}

The spin-reorientation transition is seen at 1\,T  in Fig. \ref{rhopt0} as a peak in $\mathrm{d}\rho/\mathrm{d}T$  coincident in temperature with the feature in $C_{\rm{mag}}/T$ at around 10\,K shown in Fig. \ref{Cmag}. This transition is marked with a solid arrow in Fig. \ref{rhopt0}. At each doping, the field at which the spin-reorientation transition occurs at 2.5\,K was established from the maximum of the derivative of the field dependent transverse magnetoresistance plotted in Fig. \ref{H_Pt_2K_PhaseDiag}a. This transition field is shown in Fig. \ref{H_Pt_2K_PhaseDiag}b as a function of doping. This field is in reasonable agreement with the maximum of the slope of the magnetization shown for several dopings in Fig. \ref{H_Pt_2K_PhaseDiag}c at 2.1\,K.

 It is interesting to note that in U$_2$RhC$_2$ at 2.5\,K the spin-reorientation field is only 0.5\,T. This is a remarkably small field compared to the N\'{e}el temperature of 22\,K, and it is perhaps suggestive of ferromagnetic correlations within the system. The positive Weiss constant in the modified Curie-Weiss fit to the high temperature magnetic susceptibility, discussed above, may also suggest the presence of ferromagnetic correlations. More definitive evidence for these correlations would require further investigation, however, for example by neutron scattering or NMR measurements.

\section{Superconductivity}
 Specific heat measurements indicate that superconductivity in U$_2$PtC$_2$ emerges from a renormalized, but otherwise conventional, Fermi liquid state. Fig. \ref{C_TandChi} shows that $T_c$ is maximal in the parent compound U$_2$PtC$_2$, and is suppressed as the Pt content is reduced. In contrast, the strength of the quantum critical fluctuations, as measured by the magnitude of the upturn in $C/T$, is maximal close to $x=0.7$, and is completely suppressed as Pt content increases towards x=1. This suggests that perhaps the antiferromagnetic quantum fluctuations do not enhance the superconducting pairing. Indeed, recent NMR measurements in U$_2$PtC$_2$ have shown evidence for unconventional superconductivity and more specifically spin-triplet pairing \cite{Mounce2014}. Hence, the nature of the superconducting state requires closer investigation.

 The $T_c$ of unconventional superconductors that are believed to possess a spin-triplet pairing state is very sensitive to non-magnetic impurities \cite{Analytis2006, Mackenzie1998}. Hence, the reduced $T_c$ in doped samples may arise from the pair breaking arising from additional disorder, rather than a reduction of pairing strength. In this regard, it is surprising that $T_c$ is not already completely suppressed at $x=0.9$. The residual resistivity of U$_2$PtC$_2$ is $\sim$ \SI{5}{\micro\ohm\centi\metre}. The mean free path $l$ can be estimated from the electronic specific heat coefficient per unit volume $\gamma_v=k_B^2m^*k_F/3\hbar^2$, the penetration depth $ \lambda = m^*/\mu_0 ne^2$, the residual resistivity $\rho_0=m^*/ne^2\tau$, and the Fermi wave-vector $k_F=(3\pi^2 n)^{1/3}$, where $k_B$ is the Boltzmann constant, $m^*$ is the effective electron mass, $\mu_0$ is the permeability of free space, and $\tau$ is the electron scattering time \cite{Ashcroft}. In U$_2$PtC$_2$ we estimate $l\sim$\SI{700}{\angstrom}. The intrinsic coherence length $\xi_0$ is estimated to be $\sim$\SI{70}{\angstrom}, from measurements of the upper critical magnetic field $H_{c2}$, and the equation $H_{c2}(0)=\Phi_{0}/(2\pi\xi_0^2)$, where  $\Phi_{0}$ is the flux quantum. Hence, $l\gg\xi$ and U$_2$PtC$_2$ is in the clean limit. However, the residual resistivity in $x=0.9$ is $\rho_0\sim$ \SI{42}{\micro\ohm\centi\metre}. Assuming that the effective mass and electron density are not significantly altered by the introduction of 10$\%$ Rh, this leads to a reduction of the mean free path to $l\sim$\SI{90}{\angstrom}, and is therefore comparable to the coherence length. In the extension of Abrikosov-Gor'kov theory of pair breaking scattering to unconventional superconductivity, in the limit of $\xi\approx l$ the superconductivity is predicted to be completely suppressed \cite{AbrikosovAAandGorkov1960,Larkin1965,Balian1963}. However, $T_c$ is only reduced from 1.45\,K in U$_2$PtC$_2$ to 1.09\,K in $x=0.9$. In addition, little variation of T$_c$ was seen in our measurements of U$_2$PtC$_2$, and previous measurements of U$_2$PtC$_2$ with a residual resistivity of $\sim$ \SI{10}{\micro\ohm\centi\metre}, but $T_c$ was 1.5\,K \cite{Wu1994a}. Such insensitivity to impurities is difficult to reconcile with a scenario of non-s-wave superconductivity. However, several established unconventional superconductors, such as organic \cite{Analytis2006}, cuprate \cite{Graser2007}, pnictide \cite{Fernandes2012} and heavy-fermion superconductors \cite{Smith1984,Gofryk2012}, deviate from the Abrikosov-Gor'kov theory , with explanations ranging from spatial variation of the gap function \cite{Zhitomirsky1998}, the effect of a combination of magnetic and non magnetic impurities \cite{Openov1998}, or interactions between paramagnetic impurities \cite{Galitski2002}. Hence, a definitive statement about the nature of the superconducting state will require further investigation, and in particular would benefit from the growth of single crystals and detailed studies of the gap symmetry. 

\section{Conclusion}
We have determined the phase diagram of the doping series U$_2$Pt$_x$Rh$_{(1-x)}$C$_2$, demonstrating the suppression of the antiferromagnetic phase transition in U$_2$RhC$_2$ to zero temperature close to $x=0.7$, where we observe evidence of quantum fluctuations. The antiferromagnetic state undergoes a spin-reorientation transition in an applied magnetic field. The spin-reorientation field is non-monotonic as a function of doping, with a maximum around $x=0.3$. Superconductivity is observed for $x\geq0.9$ and $T_c$ is maximal in U$_2$PtC$_2$. The suppression of $T_c$ with the increased residual resistivity of the $x=0.9$ sample is inconsistent with the extension of Abrikosov-Gor'kov theory to unconventional superconductors.

\section{Acknowledgments}
We thank J. Smith, G. Stewart, G. Meisner and Z. Fisk for useful discussions. Work at LANL was supported by the U.S. Department of Energy, Office of Science, Basic Energy Sciences, Materials Sciences and Engineering Division.

\end{document}